\documentclass[a4paper, 11pt]{jpconf}
\usepackage{graphicx}
\newcommand{\dsfrac}[2]{\displaystyle{\frac{#1}{#2}}}
\newcommand{\eqref}[1]{(\ref{#1})}

\begin{document}
\title{Numerical simulations of dynamics and emission from
  relativistic astrophysical jets}

\author{Petar Mimica}
\address{Departamento de Astronom\'ia y Astrof\'isica, Universidad de
  Valencia, 46100, Burjassot, Spain}
\ead{petar.mimica@uv.es}
\author{Miguel Angel Aloy}
\address{Departamento de Astronom\'ia y Astrof\'isica, Universidad de
  Valencia, 46100, Burjassot, Spain}
\author{Jesus M. Rueda-Becerril}
\address{Departamento de Astronom\'ia y Astrof\'isica, Universidad de
  Valencia, 46100, Burjassot, Spain}
\author{Siham Tabik}
\address{Dept. Computer Architecture, University of Málaga, 29080
  M\'{a}laga, Spain}
\author{Carmen Aloy}
\address{Departamento de Astronom\'ia y Astrof\'isica, Universidad de
  Valencia, 46100, Burjassot, Spain}

\begin{abstract}
  Broadband emission from relativistic outflows (jets) of active
  galactic nuclei (AGN) and gamma-ray bursts (GRBs) contains valuable
  information about the nature of the jet itself, and about the
  central engine which launches it. Using special relativistic
  hydrodynamics and magnetohydronamics simulations we study the
  dynamics of the jet and its interaction with the surrounding
  medium. The observational signature of the simulated jets is
  computed using a radiative transfer code developed specifically for
  the purpose of computing multi-wavelength, time-dependent,
  non-thermal emission from astrophysical plasmas. We present results
  of a series of long-term projects devoted to understanding the
  dynamics and emission of jets in parsec-scale AGN jets, blazars and
  the afterglow phase of the GRBs.
\end{abstract}

\section{Introduction}
\label{sec:introduction}

Relativistic jets are found in a variety of astrophysical scenarios
such as active galactic nuclei (AGN), microquasars, gamma-ray bursts
(GRBs) and those resulting from tidal-disruption events (TDEs). While
jet temporal and spatial timescales can vary from relatively small and
short-lived (GRB jets) to very large and long-lived (AGN jets), what
they have in common is that they are relativistic, seem to be very
collimated and appear to be launched from a region very close to the
accreting black hole (see e.g., \cite{Boettcher:2012} for a recent
review and introduction to relativistic jets).

In this paper we review a series of projects whose aim is to
understand the dynamics and emission from relativistic jets by
focusing on specific scenarios appearing in relativistic jets of AGNs,
blazars and GRBs. We begin by describing the numerical method in
section~\ref{sec:numerical}, where we describe the equations of
special relativistic magnetohydrodynamics and the code
\emph{MRGENESIS} (section~\ref{sec:RMHD}),
and then give some details about the treatment of non-thermal
particles (section~\ref{sec:nonth}) and the radiative transfer
(section~\ref{sec:radtrans}), both of which are handled by the
\emph{SPEV}. In section~\ref{sec:applications} we present three of the
applications of \emph{MRGENESIS}+\emph{SPEV} method, and we conclude
in the section~\ref{sec:conclusions}.

\section{Numerical method}
\label{sec:numerical}

The numerical method we use consists of three parts, each dealing with
a different aspect of the jet dynamics and emission. In the
section~\ref{sec:RMHD} we discuss the equations of the special
relativistic magnetohydrodynamics (RMHD). In sections~\ref{sec:nonth}
and \ref{sec:radtrans} the treatment of non-thermal particles and
emission, as well as that of the radiative transport is
explained. Finally, the structure of the \emph{SPEV} code is explained
in \ref{sec:SPEV}.

\subsection{Special relativistic magnetohydrodynamics}
\label{sec:RMHD}
We solve the system of equations of ideal special
RMHD\cite{Anile:1989}, which consists of the mass and the total-energy momentum
conservation\footnote{In this section we assume a system of units
  where the speed of light $c=1$ and the factor $1/\sqrt{4\pi}$ is
  included in the definition of the magnetic field.}
\begin{equation}\label{eq:massenmomconserve}
  \nabla_{\alpha}\left(\rho u^\alpha\right) = 0\ \ \ ;\ \ \    \nabla_{\alpha}T^{\alpha\beta} = 0\, ,
\end{equation}
where the covariant derivative is denoted by $\nabla_{\alpha}$. $\rho$
is the fluid rest-mass, $u^\alpha$ is the fluid 4-velocity, and the
energy-momentum tensor is defined as
\begin{equation}\label{eq:enmomtensor}
T^{\alpha\beta} := \rho h^* u^\alpha u^\beta + p^* \eta^{\alpha\beta} -
b^\alpha b^\beta\, .
\end{equation}
The Minkowski metric is $\eta^{\alpha\beta} = diag(-1, +1, +1, +1)$,
and the total pressure is $p^* := p + |b|^2/2$, where $p$ is the fluid
pressure. $b^\alpha$ is the magnetic field four-vector,
\begin{equation}\label{eq:b4}
  b^0:= \Gamma\left(\mathbf{v}\cdot\mathbf{B}\right)\, ,\ \ b^i:=\dsfrac{B^i}{\Gamma}
  + v^i b^0\, .
\end{equation}
Here $\mathbf{B}$ is the magnetic field 3-vector and $\mathbf{v}$ the
fluid velocity 3-vector, whose Lorentz factor is defined as
$\Gamma:=\left(1 - v^2\right)^{-1/2}$. The total specific enthalpy is
defined as
\begin{equation}\label{eq:enth}
  h^* := h + \dsfrac{\left|b\right|^2}{\rho} = 1 + \epsilon +
\dsfrac{p}{\rho} + \dsfrac{\left|b\right|^2}{\rho}\, ,
\end{equation}
where $\epsilon$ is the specific internal energy, which depends on the
equation of state $p=p(\epsilon, \rho)$. We use the TM approximation
to the Synge equation of state, where the specific
enthalpy $h$ is defined as \cite{Mimica:2010} $ h = (5/2)(p/\rho) +
\sqrt{(9/4)(p/\rho)^2 + 1}$. Finally, the magnetic field has to satisfy the induction equation
\begin{equation}\label{eq:induction}
\dsfrac{\partial \mathbf{B}}{\partial t} -
\nabla\times\left(\mathbf{v}\times\mathbf{B}\right) = 0\, ,
\end{equation}
and the divergence constraint $\nabla\cdot\mathbf{B} = 0$.

We use the code \emph{MRGENESIS}
\cite{Aloy:1999,Leismann:2005,Mimica:2007,Mimica:2009} to numerically
solve the system of equations of RMHD. \emph{MRGENESIS} uses
operator-splitting, finite-volumes, explicit method and employs
approximate Riemann solvers for the computation of intercell fluxes,
as well as total variation diminishing second and third-order
Runge-Kutta methods for the time integration (for a recent overview of
jet simulation codes see e.g. \cite{Aloy:2012}). \emph{MRGENESIS} has
been massively parallelized using OpenMP and MPI libraries and has
achieved reasonable scaling up to $10^4$ cores on the
\emph{MareNostrum} machine at the Barcelona Supercomputing Centre.

\subsection{Non-thermal particles and emission}
\label{sec:nonth}

As discussed in the section~\ref{sec:introduction}, a small fraction
of very energetic particles (``non-thermal electrons'' in the
following) is responsible for the observed jet emission. In
\cite{Mimica:2009b} we discuss in detail the method which deals with
the spatial and temporal evolution of the non-thermal electrons in a
magnetized relativistic fluid. Here we summarize the most important
properties of the method.

\subsubsection{Non-thermal electron evolution}

We assume that electrons are injected at relativistic shocks and use
the phenomenological injection term which denotes the number of
particles injected per unit volume and unit time,
\begin{equation}\label{eq:inj}
Q(\gamma) = Q_0 \left(\dsfrac{\gamma}{\gamma_{\rm min}}\right)^{-q}\
S\left(\gamma;\ \gamma_{\rm min}, \gamma_{\rm max}\right)\, ,
\end{equation}
where $\gamma$ is the electron Lorentz factor $\gamma = E / (m_e
c^2)$, $E$ being its energy and $m_e$ being its mass. $Q_0$ is the
normalization of the injection energy spectrum, $q$ is the power-law
index and the function $S(x;\ a, b)$ has the value $1$ if $a\leq x\leq
b$ and $0$ otherwise\footnote{We fix $q=2.3$ and determine the rest of
  the parameters by assuming a proportionality between the number and
  energy density of the non-thermal electrons and those of the thermal
  fluid (see equations 36 and 37 of \cite{Mimica:2009b}).}. The
temporal evolution of the electron energy distribution in the fluid
comoving frame\footnote{We assume that non-thermal electrons are
  advected with the thermal fluid. The spatial distribution of
  non-thermal electrons is represented by Lagrangian tracer
  particles.} is governed by the kinetic equation
\cite{Kardashev:1962}:
\begin{equation}\label{eq:kinetic}
  \dsfrac{\partial n(t; \gamma)}{\partial t} +
  \dsfrac{\partial}{\partial
    \gamma}\left[\dot{\gamma}n(t;\gamma)\right] = Q(\gamma)\, ,
\end{equation}
where $n(t; \gamma)$ is the number density of electrons in an interval
$\mathrm{d}\gamma$ around the Lorentz factor $\gamma$ at the time
$t$. Energy gains and losses are described by the term
$\dot{\gamma}:=\mathrm{d}\gamma / \mathrm{d}t$. For sufficiently small
intervals of time we can write the energy losses as
\begin{equation}\label{eq:losses}
\dsfrac{\mathrm{d}\gamma}{\mathrm{d}t} = k_a \gamma - k_s \gamma^2\, ,
\end{equation}
where the adiabatic gains or losses are described by the quantity
\[ k_a := \dsfrac{\mathrm{d}\ln\rho}{\mathrm{d}t}\, , \]
which denotes the compression or expansion of the fluid element, while
the synchrotron losses are denoted by
\[ k_s := -\dsfrac{4\sigma_T B'^2}{3m_e^2 c^2} \] where $B'$ is the
magnetic field strength in the fluid comoving frame and $\sigma_T$ is
the Thomson cross section.

As described in section 3.2 of \cite{Mimica:2009b}, we discretize the
electron energy space in a number of logarithmically spaced bins and
use a semi-analytic solver to accurately solve the equation
\eqref{eq:kinetic} with a minimum computational effort. 

\subsubsection{Non-thermal radiation processes}
\label{sec:radproc}

We consider the following radiation processes in our simulations:
synchrotron emission, synchrotron-self Compton (SSC) scattering, and
the external inverse-Compton (EIC) scattering.

{\bf Synchrotron emission} is produced when high-energy electrons
gyrate around the magnetic field lines, producing a non-thermal,
broadband emission spectrum. It is computed for each energy bin of the
electron distribution using the interpolation method described in the
sections 2.1.3 and 4.3.1 of \cite{Mimica:2004}.

{\bf Inverse-Compton scattering} is a scattering of a low-frequency
photon off a high-energy electron, whereby the frequency of the
outgoing photon can be many orders of magnitude larger than that of
the incoming one. In our numerical method it is computed analytically
(see section 2.2.2 of \cite{Mimica:2004}) assuming that the incoming
emission spectrum is a power-law, as is the energy spectrum of the
electrons off which the photons scatter. This method has been used in
\cite{Mimica:2011} to compute the EIC scattering of the ultraviolet
stellar, photons, as well as in \cite{Mimica:2012} to compute both the
SSC and EIC emission in the blazar jets.

\subsection{Radiative transfer}
\label{sec:radtrans}

To compute the time- and frequency-dependent synthetic image we use
the algorithm described in the Appendix A of \cite{Mimica:2009b}. Both
this algorithm and the one described in the previous section are
building blocks of the code \emph{SPEV}. \emph{SPEV} solves the
radiative transfer equation
\begin{equation}\label{eq:radtrans}
  \dsfrac{\mathrm{d}I_\nu}{\mathrm{d}s} = j_\nu - \alpha_\nu I_\nu\, ,
\end{equation}
where $I_\nu$, $j_\nu$ and $\alpha_\nu$ are the specific intensity,
emissivity and absorption coefficient at frequency $\nu$. $s$ is the
distance along the line of sight which is seen  by the observer at
the time $T$. The relation between the time of observation and the
time $t$ in the laboratory frame (the frame in which the jet evolution is
simulated) is given by
\begin{equation}\label{eq:tobs}
t = T + (D - s)/c\ , 
\end{equation}
where $D$ is the distance of the jet injection point from the Earth
and we choose that $s = 0$ at that point. As can be seen, in order to
solve the equation \eqref{eq:radtrans}, it is not enough to process
each saved simulation snapshot at time $t_{\rm snapshot}$. Instead, it
is necessary to process \emph{all} snapshots \emph{simultaneously}
because of the fact that distance from the observer and the time at
which the snapshot has been taken are not independent, but must
satisfy the equation \eqref{eq:tobs}. In practice, this means that the
radiative transfer equation \eqref{eq:radtrans} cannot be solved
before \emph{all} the information along the line of sight has been
gathered. This makes the problem tightly coupled and highly non-local,
complicating the code paralellization and increasing the CPU and
memory costs. 

\subsubsection{Structure and paralellization of the \emph{SPEV} code}
\label{sec:SPEV}

The evolution and emission from the non-thermal particles
(section~\ref{sec:nonth}) and the computation of the resulting
emission (section~\ref{sec:radtrans}) are done by the \emph{SPEV} code
whose structure can be sketched as follows: after creating the initial
Lagrangian particles (representing the pre-existing non-thermal
electrons), \emph{SPEV} iteratively runs through the three phases:
evolution of existing particles, injection of new particles, and
calculation of the emission (i.e., contribution to the pixels of a
virtual detector on which the emission is computed).

\emph{SPEV} has two important characteristics very useful for its
paralellization: the evolution of each Lagrangian particle during the
simulation does not affect the evolution of the rest of the particles,
and the contribution to each pixel of the detector is independent from
the contribution to any other pixel. Therefore, \emph{SPEV} is
parallelized over particles \emph{and} detector pixels. The data is
partitioned into different sets and distributed among different
computing units (e.g., multiple multicore-nodes connected by means of
the network or by means of buses). Each set of particles can evolve
separately without the need of any kind of communications during the
iterative process. When the iterative process ends, data is
transferred to the master node and then the radiative transfer
equation \eqref{eq:radtrans} is solved for each pixel. More details
about the parallel implementation are provided in
\cite{Tabik:2011,Tabik:2012}. This parallel approach has shown very
good scalability on different HPC-systems, especially on medium scale
multi-sockets multicores of up to 50 cores.

\section{Applications}
\label{sec:applications}

In this section we outline a number of application of \emph{MRGENESIS}
+ \emph{SPEV} approach to relativistic jets. In
section~\ref{sec:pcjets} we discuss the radio emission from
parsec-scale jets. The section~\ref{sec:blazars} is devoted to address
the highly variable and highly energetic emission from the blazars, while
section~\ref{sec:GRBs} presents results of a long-term study of
gamma-ray burst afterglows.

\subsection{Parsec-scale jets}
\label{sec:pcjets}

\begin{figure}[h]
\begin{minipage}{18pc}
\includegraphics[width=18pc]{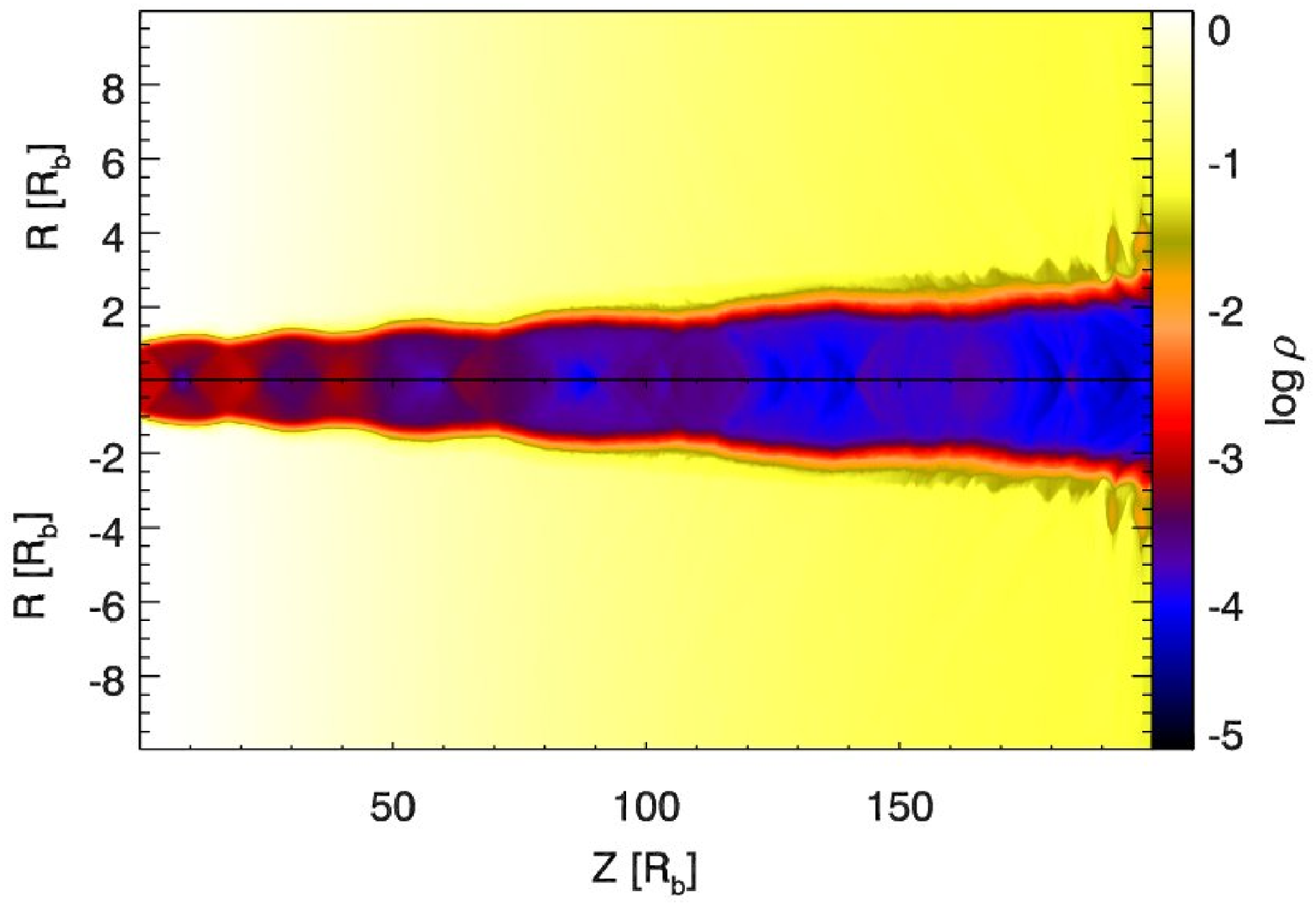}
\caption{\label{fig:jet} Logarithm of the density of a steady-state
  jet propagating through a stratified atmosphere. The jet is
  initially under-dense and over-pressured with respect to the
  surrounding medium. The density is shown in arbitrary units, while
  the length scale is in the units of the jet radius at the left grid
  boundary. See section~2 of \cite{Mimica:2009b} for the details of
  the hydrodynamic models. Note that the vertical scale is enlarged
  for the better visualization of the jet.}
\end{minipage}\hspace{2pc}%
\begin{minipage}{18pc}
\includegraphics[width=14pc]{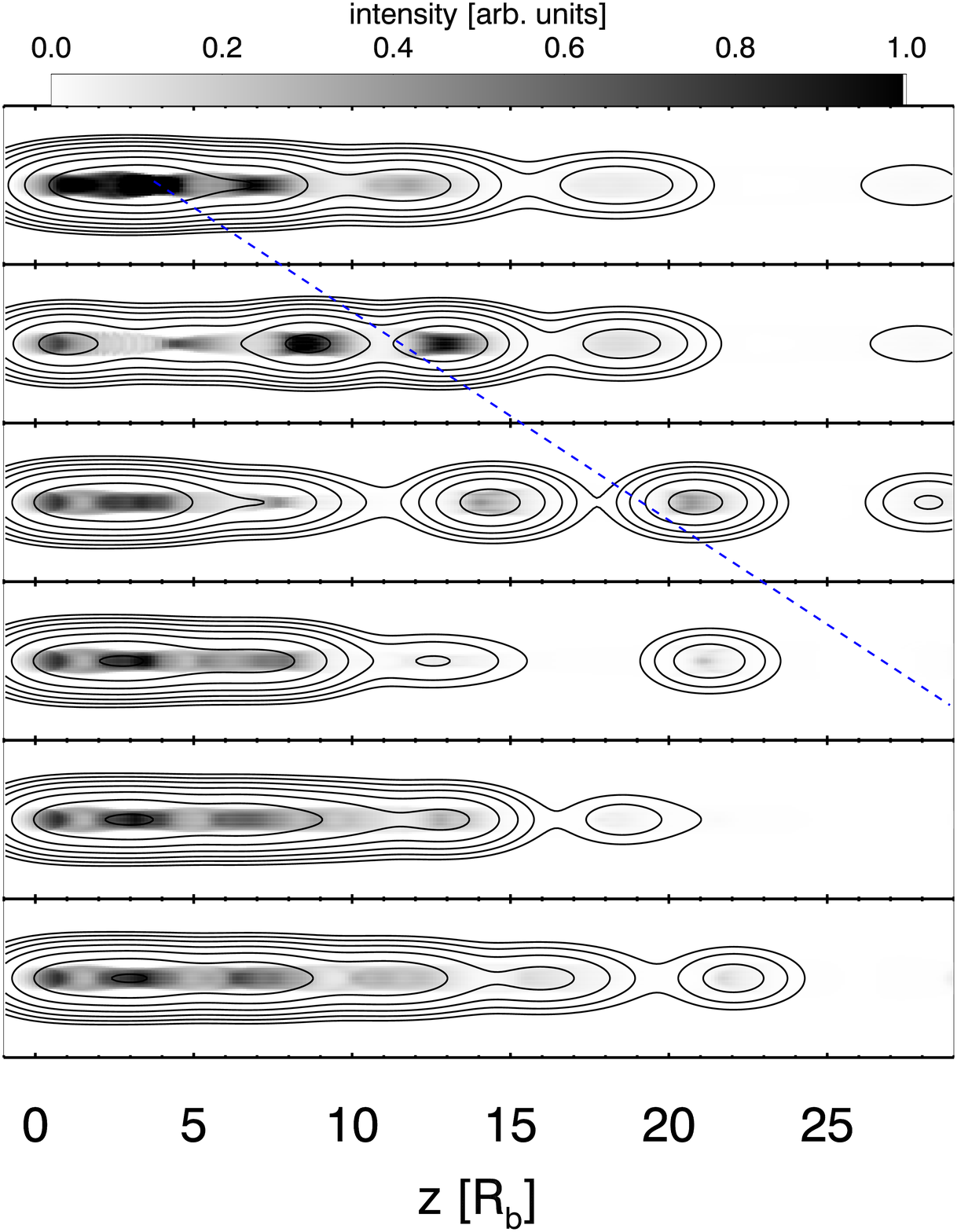}
\caption{\label{fig:jetemiss} Radio emission at $43$ GHz resulting
  from a temporary velocity perturbation propagating through the
  steady-state jet shown on Figure~\ref{fig:jet}. The gray shades in
  the panels show the radiation intensity (in arbitrary units) at
  (from top to bottom) $0.02$, $0.39$, $0.75$, $1.12$, $1.94$ and
  $4.58$ years since the injection of the perturbation. The contours
  shows the image degraded to the resolution available with today's
  VLBI technique. The dashed blue line shows the approximate
  trajectory of the main component. See section~7 of
  \cite{Mimica:2009b} for more details.}
\end{minipage}
\end{figure}

Parsec-scale jets are radio features on a scale of several light
years. They are seen emerging from AGNs over periods of months to
years. It is widely accepted that the radio maps are not direct
observations of the physical quantities (density, pressure, velocity,
composition, etc.) in the jet, but are influenced by a number of
relativistic effects (Doppler shift, beaming) as well as by the
degradation of the image due to a finite resolution of the radio
telescopes. By using numerical simulations of the jet dynamics and,
subsequently, by computing the synthetic radio maps we can test
theoretical jet models and directly compare them to the observations,
as well as study some of the events which are expected to occur in
them\footnote{Our work on blazar jets (section~\ref{sec:blazars}) as
  well as studies of the stability of initially very magnetized
  ($\sigma\geq 100$) jets (e.g. \cite{Giannios:2006}) indicate that at
  parsec-scales distances AGN jets are expected to be at most very
  weakly magnetized (e.g. $\sigma \leq 10^{-4}$)}. In a previous work
\cite{Mimica:2009b} we studied one such event, which is an injection
of a perturbation into a steady jet shown on
Figure~\ref{fig:jet}. This jet is over-pressured and under-dense with
respect to the surrounding medium and has a number of recollimation
shocks, which can be seen as knots in the density profile shown on
Figure~\ref{fig:jet}.

The perturbation has a form of a temporary increase in the fluid
velocity at the jet injection point located at the left grid
boundary. This results in the development of shock and rarefaction
waves which propagate down the jet and interact with the recollimation
shocks, resulting in a complex, non-linear evolution of the
perturbation and several trailing components (see section~7 of
\cite{Mimica:2009b} or more details). Figure~\ref{fig:jetemiss} shows
the radio maps taken at six different epochs of the jet evolution. In
the first three panels we can see the main component (perturbation) as
a dark region propagating through the jet. In the last three panels we
see a gradual reestablishment of the steady jet.

The contours show the image degraded to the resolution available with
today's radio telescopes. As can be appreciated, many of the
small-scale details of the jet evolution and emission are lost in the
relatively strong blurring of the image due to the low resolution of
the actually observable radio maps. Therefore it is important to use
the hydrodynamic simulations and possess the ability to compute
time-dependent radio maps from those simulations because this enables
us to compute both the full-resolution and the degraded images and
thus evaluate which of the observed features of the jet dynamics and
emission are intrinsic to the jet and which are the artefact of the
\emph{finite} observational resolution.

\subsection{Blazar jets}
\label{sec:blazars}

\begin{figure}[h]
\begin{minipage}{18pc}
\includegraphics[width=18pc]{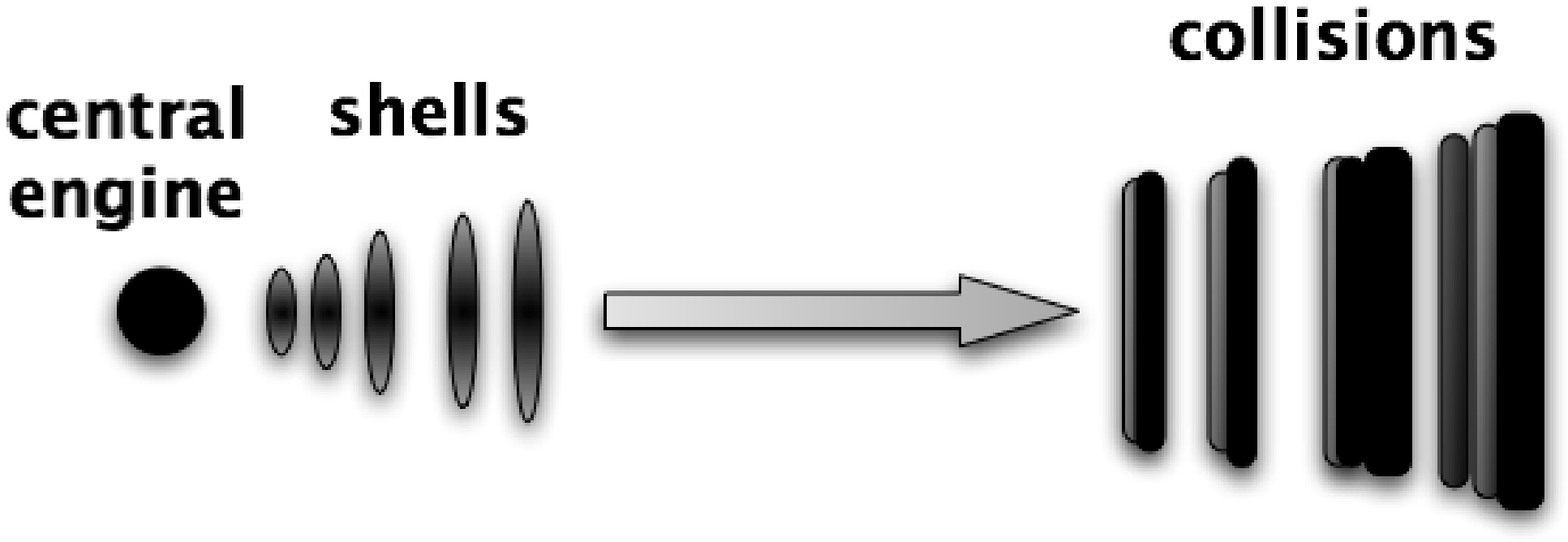}
\caption{\label{fig:insh} An illustration of the internal shocks
  model: an intermittently active central engine (supermassive black
  hole at the center of the galaxy) ejects inhomogeneous shells which
  collide and produce observed flares.}
\end{minipage}\hspace{2pc}%
\begin{minipage}{18pc}
\includegraphics[width=18pc]{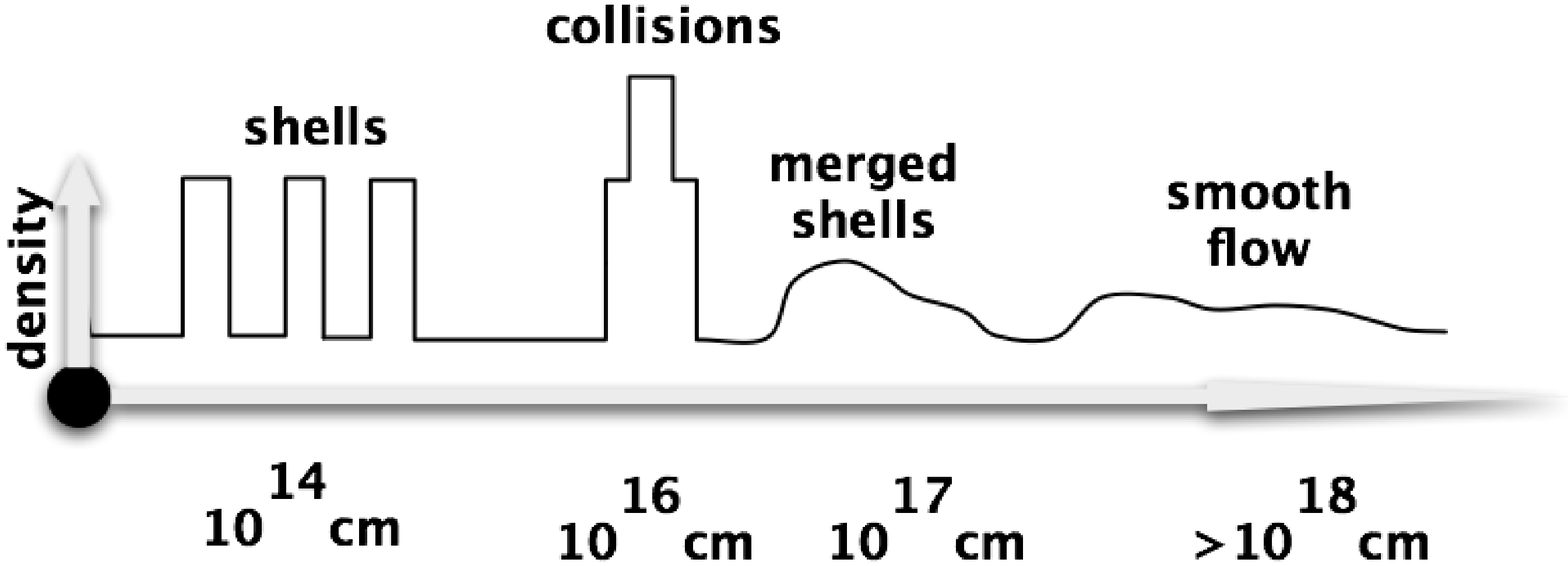}
\caption{\label{fig:insh2} Distance scales in the internal shocks
  model in blazars: the shells, whose characteristic size is $\approx
  10^{14}$ cm, collide at distances $\approx 10^{16} - 10^{17}$ cm
  from the central engine, merge and continue flowing as a smooth jet
  at parsec-scale distances ($\approx 10^{18}$ cm).}
\end{minipage}
\end{figure}

Blazars are jets pointing almost directly towards us. They are
characterised by high variability of their emission, often in the form
of flares observed by X-ray satellites\footnote{For an example of
  X-ray flares from a well known blazar Mrk 421 see e.g.,
  \cite{Brinkmann:2005}}. It is thought that the cause of the flares
are the collisions of parts of the jet with different velocities
(``shells'' in the following), whereby a fraction of the shell kinetic
energy is dissipated by the shocks which propagate through the shells
as a result of their collision, and a fraction of the dissipated
energy is radiated and observed as a flare. This is called the
internal shock scenario
\cite{Rees:1994,Kobayashi:1997,Daigne:1998,Mimica:2004b,Kino:2004,Mimica:2005,Mimica:2007,Bosnjak:2009,Mimica:2010,Mimica:2012}. Figures~\ref{fig:insh}
and \ref{fig:insh2} illustrate the internal shocks model and the
typical distance scales at which shells collide in blazar jets.

We study the dynamics and emission from single shell collisions, using
one-dimensional \cite{Mimica:2005} and two-dimensional
\cite{Mimica:2004b} hydrodynamic simulations, as well as
one-dimensional magnetohydrodynamic simulations \cite{Mimica:2007}. We
found that the lateral shell spreading is negligible on the length
scales considered and that the one-dimensional approximation is
accurate \cite{Mimica:2004b}. Furthermore, we find that the
magnetization of the flow plays a crucial role in the efficiency of
the conversion of kinetic into thermal energy (and subsequently into
radiation). From the computational point of view we found the
hydrodynamic simulations to be rather expensive and unsuitable for
parameter space studies.

In the follow-up works \cite{Mimica:2010,Mimica:2012} we simplify the
hydrodynamic approach in order to be able to devote more resources to
the computation of the emission and to be able to cover larger
parameter space. We study a large number of shell collisions where
both shells have the same energy and size, but their velocity and the
degree of magnetization can vary. We denote the shell Lorentz factor
by $\Gamma:=(1 - v^2/c^2)^{-1/2}$, where $v$ is the shell
velocity. The magnetization for cold and relativistic shells
(e.g. $p/\rho c^2\ll 1$ and $\Gamma\gg 1$, where $\rho$ and $p$ are
the shell rest-mass density and the thermal pressure) is defined as
$\sigma:=B^2/(4\pi \Gamma^2 \rho c^2)$ . We compute the exact
solution of the Riemann problem\footnote{We use the exact Riemann
  solver \cite{Romero:2005} for the one-dimensional ideal relativistic
  magnetohydrodynamic Riemann problem with magnetic fields
  perpendicular to the flow velocity.} for each shell collision and
determine the strength and velocity of the shocks which are formed by
the shell interaction. By covering a large parameter space of possible
shell magnetization (we studied $10^6$ shell collisions with $\sigma$
ranging from $10^{-6}$ to $10^3$ for each shell independently) we find
that the ``sweet spot'' for the efficiency of the kinetic energy
conversion is not for $\sigma \ll 1$, as might be expected (the
smaller the $\sigma$, the easier it is to shock the fluid), but rather
for $(\sigma_L = 1,\ \sigma_R = 0.1)$, where subscripts $L$ and $R$
denote the faster (left) and the slower (right) shell, respectively
\cite{Mimica:2010}. This result makes the magnetized internal shock
model viable. In the follow-up work \cite{Mimica:2012} we compute the
multi-wavelength, time-dependent emission from $\approx 10^3$ shell
collisions.\footnote{\label{foot:comp} We could not compute the
  emission from all of the $10^6$ shell collisions studied in
  \cite{Mimica:2010} because it was not feasible with present
  supercomputers (the computation time for the emission from $10^3$
  shell collisions exceeds $200$ thousand hours).}

\begin{figure}[h]
\begin{minipage}{18pc}
\includegraphics[width=18pc]{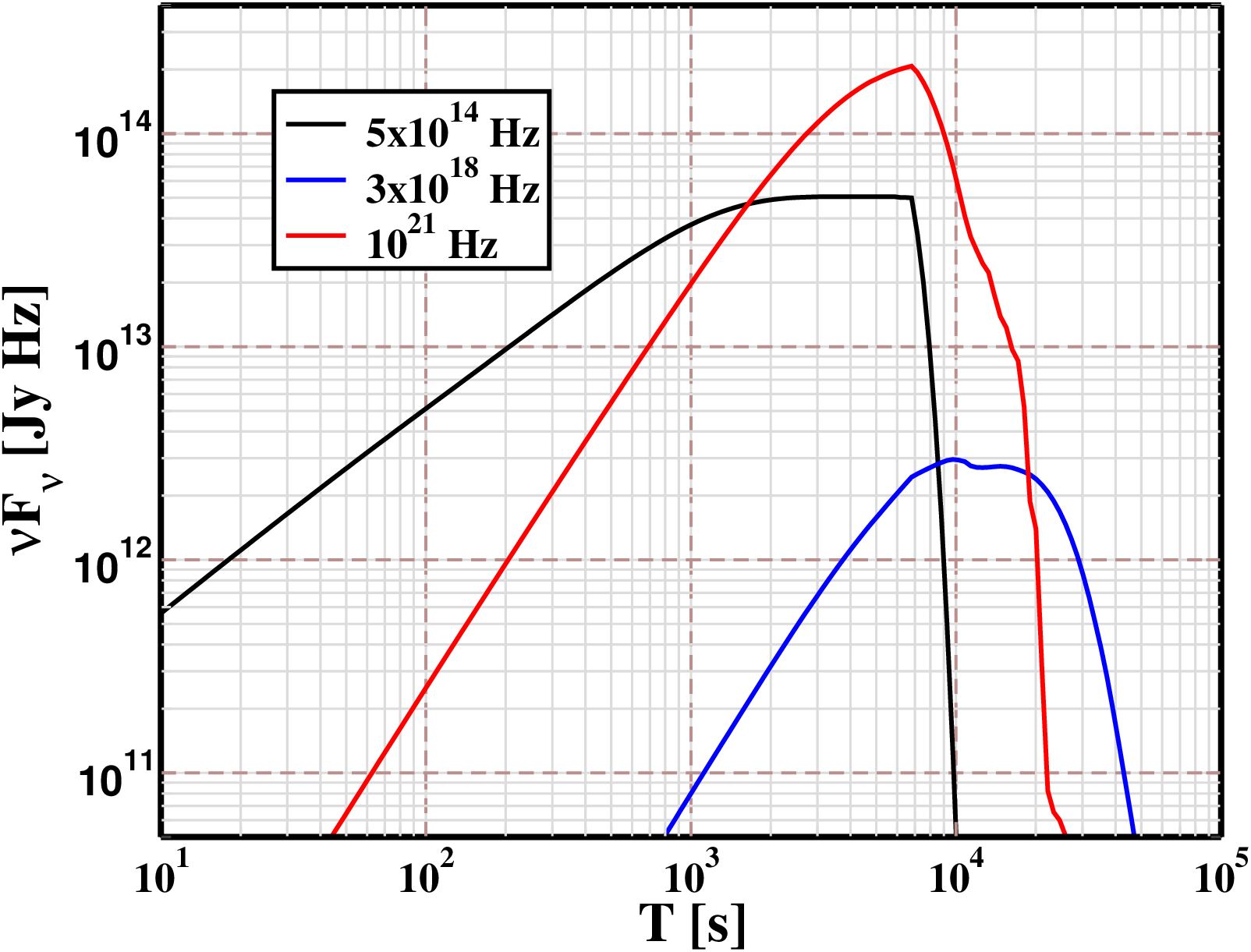}
\caption{\label{fig:bllc} Optical ($5\times 10^{14}$ Hz), X-ray
  ($3\times 10^{18}$ Hz) and $\gamma$-ray ($10^{21}$ Hz) flare (black,
  blue and red lines) produced by a collision of a
  faster shell with the initial Lorentz factor $\Gamma_L = 20$ with a
  slower shell $\Gamma_R = 10$. The shell magnetization is $\sigma_L =
  10^{-3}$ and $\sigma_R = 10^{-1}$. The shell geometry is cylindrical
  (radius $3\times 10^{14}$ m, length $6\times 10^{11}$ m) and their
  initial energy $10^{52}$ erg. The flux is shown in units Jy Hz ($1$
  Jy Hz $= 10^{-23}$ W / m$^2$).}
\end{minipage}\hspace{2pc}%
\begin{minipage}{18pc}
\includegraphics[width=18pc]{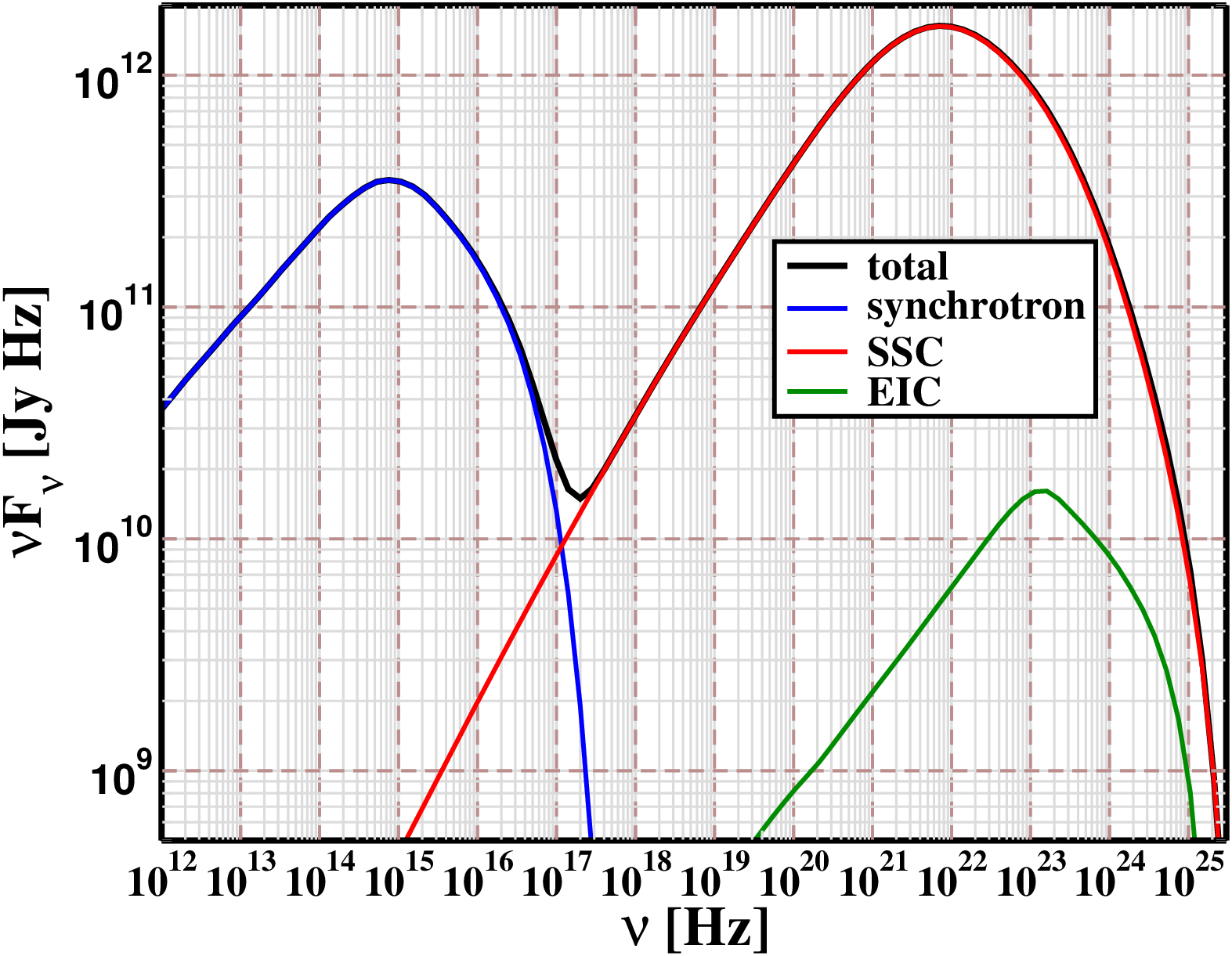}
\caption{\label{fig:blspec} Average spectrum of the flare shown in
  figure~\ref{fig:bllc}. The total spectrum is shown in black, while
  the synchrotron, SSC and EIC contributions are shown in blue, red
  and green, respectively. The seed photons for the EIC are assumed to
  be monochromatic (frequency $2\times 10^{14}$ Hz). The viewing angle
  (angle between the jet axis and the line of sight) is $5^\circ$, and the
  jet is assumed to be located at redshift $z=0.5$.}
\end{minipage}
\end{figure}

Figures~\ref{fig:bllc} and \ref{fig:blspec} show an example of the
light curve and spectra produced by a collision of two moderately
magnetized shells in a typical blazar jet (Rueda, Mimica \& Aloy,
2012, in preparation). In figure~\ref{fig:bllc} can see that the
optical and $\gamma$-ray light curves peak somewhat earlier than the
X-ray light curve. The reason is that the SSC emission
(section~\ref{sec:radproc}) is produced by the scattering of the
synchrotron emission which needs a finite time to travel from the
place where it is produced to the site of its scattering. The sharp
drop in the emission is related to the moment when shocks cross the
shells and cease dissipating energy. Average spectra are shown in
figure~\ref{fig:blspec}. The low-frequency emission is dominated by
the synchrotron emission, while the high-energy spectrum is dominated
by the SSC component. As can be seen, the correct computation of
the SSC component is essential, and since this component is
computationally most expensive, it puts an upper limit on a number of
models which can be computed on today's machines (see
footnote~\ref{foot:comp}).

\subsection{Gamma-ray burst afterglows}
\label{sec:GRBs}

Gamma-ray bursts (GRBs) are produced by an internal dissipation of
energy in the relativistic outflow. The GRB afterglow emission is
produced by the interaction of the relativistic jet with the external
(circumburst) medium. It is not known to what extent the jet is
magnetized. If the outflow is initially dominated by the thermal
energy (fireball model \cite{Goodman:1986,Paczynski:1986}) then it is
expected that its $\sigma\ll 1$. If, on the other hand, the outflow is
initially dominated by the Poynting-flux
\cite{Usov:1992,Meszaros:1997}, then it is possible to find a
significantly magnetized jet at the onset of the afterglow phase.

\begin{figure}[h]
\begin{minipage}{18pc}
\includegraphics[width=18pc]{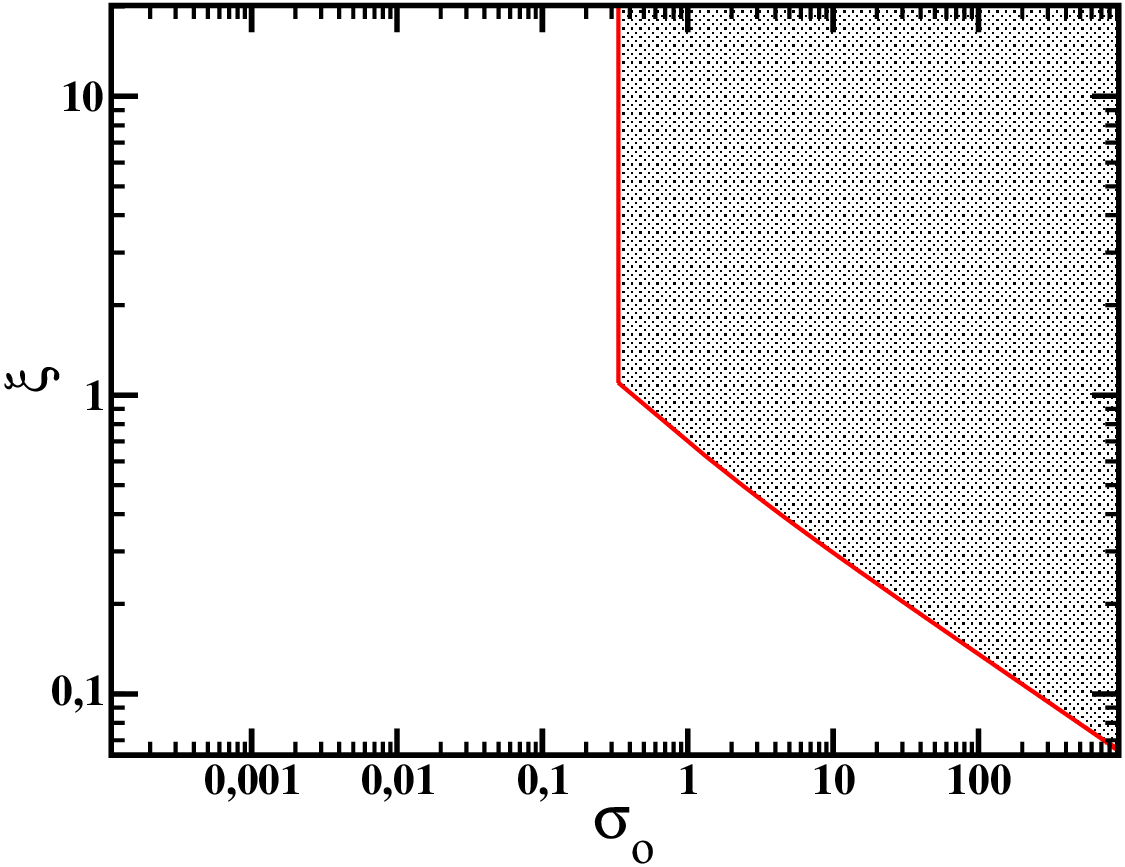}
\caption{\label{fig:xisig} Parameter space of the ejecta
  magnetization $\sigma_0$ and $\xi:= \left(3E / 4\pi
    \rho_{\rm ext}c^2\right)^{1/6} \Delta_0^{-1/2} \Gamma_0^{-4/3}$,
  where $E$, $\Delta_0$ and $\Gamma_0$ are the ejecta initial energy, width
  and the Lorentz factor, and $\rho_{\rm ext}$ is the
  ext. medium density. The shaded region shows the ejecta which do
  not form a reverse shock, while the vertical boundary at $\sigma_0=0.3$ indicates
  those ejecta whose radial spreading becomes significant and makes
  the formation of a reverse shock inevitable \cite{Giannios:2008}.}
\end{minipage}\hspace{2pc}%
\begin{minipage}{18pc}
\includegraphics[width=18pc]{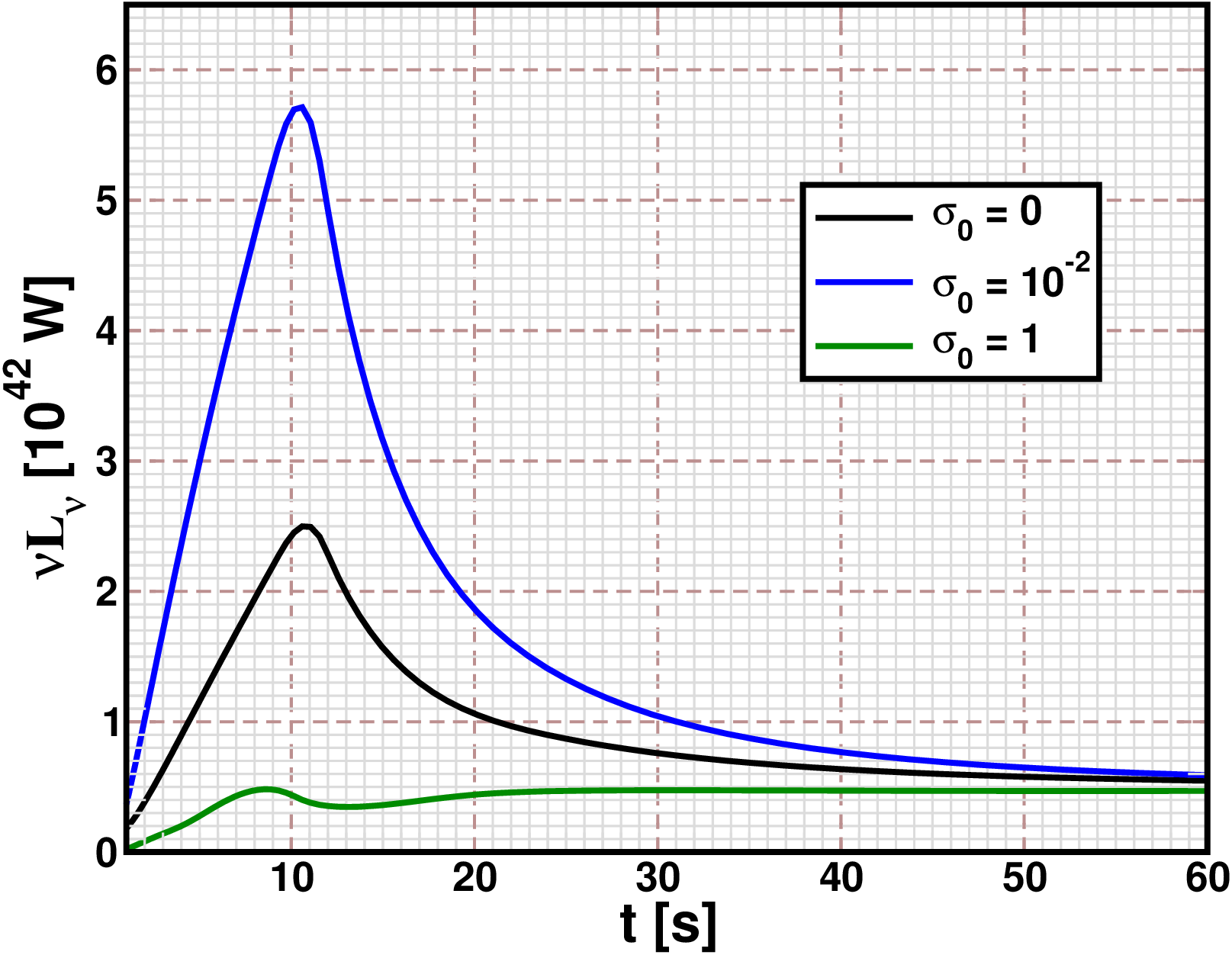}
\caption{\label{fig:R-band} Optical ($5\times 10^{14}$ Hz) light curve
  produced by the non-magnetized ($\sigma_0 = 0$), weakly magnetized
  ($\sigma_0=0.01$ and strongly-magnetized ($\sigma_0=1$) shells
  (black, blue and green lines, respectively) whose initial conditions
  are $E = 3\times 10^{47}$ J, $\Gamma_0 = 300$ and $\Delta_0 = 3
  \times 10^{9}$ m. The non/weakly magnetized ejecta have a strong
  reverse shock which produces an optical flash, while the strongly
  magnetized ejecta light curve does not have this feature \cite{Mimica:2010b}.}
\end{minipage}
\end{figure}

We study the interaction of an arbitrarily magnetized GRB jet with the
external medium. We model the interaction by considering the
deceleration of a dense, magnetized shell as it moves through and
compresses a constant density medium in front of it. One of the
crucial differences between the evolution of a non-magnetized and a
magnetized shell is in the existence of a reverse shock (a shock
propagating through the shell): if $\sigma$ is high enough, then the
interaction between the shell and the external medium is not strong
enough to shock the shell material itself. By analyzing timescales of
the shell deceleration and the propagation of magnetohydrodynamic
waves through the shell we find the conditions for the existence of a
reverse shock depending on the GRB magnetization, energy, Lorentz
factor, duration and the density of the external medium
\cite{Giannios:2008}. Figure~\ref{fig:xisig} shows the parameter space
spanned by the ejecta magnetization and a dimensionless parameter
$\xi$ (see figure caption for definition). As we can see, there is a
substantial portion of the parameter space (shaded region), where the
reverse shock does not form. Since it is absent, the particles are not
accelerated and we expect to fail to observe an emission from this
shock during the early afterglow, which would be consistent with the
absence of such observations in most GRB optical afterglows
\cite{Gomboc:2009}.

In the follow-up works \cite{Mimica:2009,Mimica:2010b} we confirmed
these results by means of numerical simulations. Using
\emph{MRGENESIS} we performed high-resolution, long-term
one-dimensional relativistic magnetohydrodynamic simulation of a
shell-medium interaction, and then use \emph{SPEV} to compute the
resulting light curves. Figure~\ref{fig:R-band} shows an example of
the optical light curve produced by non-magnetized, weakly-magnetized
and strongly-magnetizeed shells with the same initial energy, Lorentz
factor and width. As can be seen, the strongly magnetized shell light
curve does not possess the optical flash feature that the other two
possess. Therefore, strong magnetization might be required for the
paucity of the observed optical flashes to be explained. Since most of
the observed GRBs have $\xi$ in the range $[0.3, 3]$, this would
require $\sigma > 0.1$ for almost all GRBs
\cite{Giannios:2008,Mimica:2010b}. This would imply that GRB jets
could be strongly magnetized even at such late stages of their
evolution as the afterglow phase, which puts constraints on the amount
of magnetic dissipation that can happen at previous stage (prompt
emission).

The simulations needed to correctly compute afterglow light curves are
very computationally demanding due to the high resolution required in the
direction of the jet motion (e.g. more than $10^4$ zones in the
initial shell) and the long timescales needed to compute the light
curve of sufficient duration (approximately $10^7$ numerical
iterations). Therefore, performing this type of calculations to cover
a full parameter space of feasible GRB afterglows is not possible at
the moment. However, the existence of re-scaling laws (see section~4.4
of \cite{Mimica:2009}) can help us to perform a small number of
simulations and then extend their results to cover a somewhat larger
parameter space.

\section{Conclusions}
\label{sec:conclusions}

We have presented a framework consisting of a numerical relativistic
magnetohydrodynamic code \emph{MRGENESIS} and the non-thermal
radiative transfer code \emph{SPEV}. We show how its modular nature
has enabled it to be successfully been applied to a number of current
problems and issues in the relativistic jets physics, and how it
enables us, via the computation of synthetic observations, to directly
compare the results of our simulations with the observations from
ground- and space-based telescopes. In the future we plan to improve
the scaling of \emph{MRGENESIS} to more than $10^4$ cores, and also to
add new radiative processes to \emph{SPEV} (e.g., polarization,
improved inverse-Compton, radiative transfer in curved space-times,
etc.). The code scaling improvements will allow us to perform full 3D
magnetohydrodynamic simulations and radiative transfer calculations,
relevant for astrophysical scenarios where axial symmetry cannot be
assumed.

\section*{Acknowledgments}
PM, MAA and CA acknowledge the support from the ERC grant CAMAP-259276
and the grants AYA2010-21097-C03-01 and PROMETEO-2009-103. JMRB
acknowledges the support from the Grisolia fellowship
GRISOLIA/2011/041. PM, MAA, ST and CA acknowledge the support from the
Consolider grant CSD2007-00050. The authors thankfully acknowledge the
computer resources, technical expertise and assistance provided by the
Barcelona Supercomputing Center - Centro Nacional de
Supercomputaci\'on.

\section*{References}
\bibliography{Mimica.bib}

\providecommand{\newblock}{}
\begin{thebibliography}{10}
\expandafter\ifx\csname url\endcsname\relax
  \def\url#1{{\tt #1}}\fi
\expandafter\ifx\csname urlprefix\endcsname\relax\def\urlprefix{URL }\fi
\providecommand{\eprint}[2][]{\url{#2}}

\bibitem{Boettcher:2012}
B\"{o}ttcher M, Harris D~E and Krawczynski H 2012 {\em Introduction and
  Historical Perspective;\ {\rm in} Relativistic Jets from Active Galactic
  Nuclei (eds M. Böttcher, D. E. Harris and H. Krawczynski)\/} (Wiley-VCH
  Verlag GmbH \& Co. KGaA) chap~1, pp 1--16 ISBN 9783527641741

\bibitem{Anile:1989}
{Anile} A~M 1989 {\em {Relativistic fluids and magneto-fluids: With
  applications in astrophysics and plasma physics}\/} (Cambridge University
  Press) ISBN 9780521304061

\bibitem{Mimica:2010}
{Mimica} P and {Aloy} M~A 2010 {\em \mnras\/} {\bf 401} 525--532
  (\textit{Preprint} \eprint{0909.1328})

\bibitem{Aloy:1999}
{Aloy} M~A, {Ib{\'a}{\~n}ez} J~M, {Mart{\'{\i}}} J~M and {M{\"u}ller} E 1999
  {\em \apjs\/} {\bf 122} 151--166 (\textit{Preprint}
  \eprint{arXiv:astro-ph/9903352})

\bibitem{Leismann:2005}
{Leismann} T, {Ant{\'o}n} L, {Aloy} M~A, {M{\"u}ller} E, {Mart{\'{\i}}} J~M,
  {Miralles} J~A and {Ib{\'a}{\~n}ez} J~M 2005 {\em \aap\/} {\bf 436} 503--526

\bibitem{Mimica:2007}
{Mimica} P, {Aloy} M~A and {M{\"u}ller} E 2007 {\em \aap\/} {\bf 466} 93--106
  (\textit{Preprint} \eprint{arXiv:astro-ph/0611765})

\bibitem{Mimica:2009}
{Mimica} P, {Giannios} D and {Aloy} M~A 2009 {\em \aap\/} {\bf 494} 879--890
  (\textit{Preprint} \eprint{0810.2961})

\bibitem{Aloy:2012}
Aloy M~A and Mimica P 2012 {\em Simulations of Jets from Active Galactic Nuclei
  and Gamma-Ray Bursts;\ {\rm in} Relativistic Jets from Active Galactic Nuclei
  (eds M. Böttcher, D. E. Harris and H. Krawczynski)\/} (Wiley-VCH Verlag GmbH
  \& Co. KGaA) chap~10, pp 297--339 ISBN 9783527641741

\bibitem{Mimica:2009b}
{Mimica} P, {Aloy} M~A, {Agudo} I, {Mart{\'{\i}}} J~M, {G{\'o}mez} J~L and
  {Miralles} J~A 2009 {\em \apj\/} {\bf 696} 1142--1163 (\textit{Preprint}
  \eprint{0811.1143})

\bibitem{Kardashev:1962}
{Kardashev} N~S 1962 {\em \sovast\/} {\bf 6} 317

\bibitem{Mimica:2004}
{Mimica} P 2004 {\em {Numerical Simulations of Blazar Jets and their
  Non-thermal Radiation}\/} Ph.D. thesis Max-Planck-Institut f{\"u}r
  Astrophysik

\bibitem{Mimica:2011}
{Mimica} P and {Giannios} D 2011 {\em \mnras\/} {\bf 418} 583--590
  (\textit{Preprint} \eprint{1106.1903})

\bibitem{Mimica:2012}
{Mimica} P and {Aloy} M~A 2012 {\em \mnras\/} {\bf 421} 2635--2647
  (\textit{Preprint} \eprint{1111.6108})

\bibitem{Tabik:2011}
{Tabik} S, {Mimica} P, {Plata} O, {Zapata} E~L and {Romero} L~F 2011 {\em HiPC
  2011\/} {\bf 1} 1--10

\bibitem{Tabik:2012}
{Tabik} S, {Romero} L~F, {Mimica} P, {Plata} O and {Zapata} E~L 2012 {\em
  Computer Physics Communications\/} {\bf 183} 1937--1946

\bibitem{Giannios:2006}
{Giannios} D and {Spruit} H~C 2006 {\em \aap\/} {\bf 450} 887--898
  (\textit{Preprint} \eprint{arXiv:astro-ph/0601172})

\bibitem{Brinkmann:2005}
{Brinkmann} W, {Papadakis} I~E, {Raeth} C, {Mimica} P and {Haberl} F 2005 {\em
  \aap\/} {\bf 443} 397--411 (\textit{Preprint}
  \eprint{arXiv:astro-ph/0508433})

\bibitem{Rees:1994}
{Rees} M~J and {Meszaros} P 1994 {\em \apjl\/} {\bf 430} L93--L96
  (\textit{Preprint} \eprint{arXiv:astro-ph/9404038})

\bibitem{Kobayashi:1997}
{Kobayashi} S, {Piran} T and {Sari} R 1997 {\em \apj\/} {\bf 490} 92
  (\textit{Preprint} \eprint{arXiv:astro-ph/9705013})

\bibitem{Daigne:1998}
{Daigne} F and {Mochkovitch} R 1998 {\em \mnras\/} {\bf 296} 275--286
  (\textit{Preprint} \eprint{arXiv:astro-ph/9801245})

\bibitem{Mimica:2004b}
{Mimica} P, {Aloy} M~A, {M{\"u}ller} E and {Brinkmann} W 2004 {\em \aap\/} {\bf
  418} 947--958 (\textit{Preprint} \eprint{arXiv:astro-ph/0401266})

\bibitem{Kino:2004}
{Kino} M, {Mizuta} A and {Yamada} S 2004 {\em \apj\/} {\bf 611} 1021--1032
  (\textit{Preprint} \eprint{arXiv:astro-ph/0404555})

\bibitem{Mimica:2005}
{Mimica} P, {Aloy} M~A, {M{\"u}ller} E and {Brinkmann} W 2005 {\em \aap\/} {\bf
  441} 103--115 (\textit{Preprint} \eprint{arXiv:astro-ph/0506636})

\bibitem{Bosnjak:2009}
{Bo{\v s}njak} {\v Z}, {Daigne} F and {Dubus} G 2009 {\em \aap\/} {\bf 498}
  677--703 (\textit{Preprint} \eprint{0811.2956})

\bibitem{Romero:2005}
{Romero} R, {Mart{\'{\i}}} J~M, {Pons} J~A, {Ib{\'a}{\~n}ez} J~M and {Miralles}
  J~A 2005 {\em Journal of Fluid Mechanics\/} {\bf 544} 323--338
  (\textit{Preprint} \eprint{arXiv:astro-ph/0506527})

\bibitem{Goodman:1986}
{Goodman} J 1986 {\em \apjl\/} {\bf 308} L47--L50

\bibitem{Paczynski:1986}
{Paczynski} B 1986 {\em \apjl\/} {\bf 308} L43--L46

\bibitem{Usov:1992}
{Usov} V~V 1992 {\em \nat\/} {\bf 357} 472--474

\bibitem{Meszaros:1997}
{Meszaros} P and {Rees} M~J 1997 {\em \apjl\/} {\bf 482} L29 (\textit{Preprint}
  \eprint{arXiv:astro-ph/9609065})

\bibitem{Giannios:2008}
{Giannios} D, {Mimica} P and {Aloy} M~A 2008 {\em \aap\/} {\bf 478} 747--753
  (\textit{Preprint} \eprint{0711.1980})

\bibitem{Gomboc:2009}
{Gomboc} A, {Kobayashi} S, {Mundell} C~G, {Guidorzi} C, {Melandri} A, {Steele}
  I~A, {Smith} R~J, {Bersier} D, {Carter} D and {Bode} M~F 2009 {Optical
  flashes, reverse shocks and magnetization} {\em American Institute of Physics
  Conference Series\/} ({\em American Institute of Physics Conference Series\/}
  vol 1133) ed {Meegan} C, {Kouveliotou} C and {Gehrels} N pp 145--150
  (\textit{Preprint} \eprint{0902.1830})

\bibitem{Mimica:2010b}
{Mimica} P, {Giannios} D and {Aloy} M~A 2010 {\em \mnras\/} {\bf 407}
  2501--2510 (\textit{Preprint} \eprint{1004.2720})

\end{thebibliography}

\end{document}